\documentclass[graybox]{svmult}

\usepackage{mathptmx}       
\usepackage{helvet}         
\usepackage{courier}        
\usepackage{type1cm}        
\usepackage{makeidx}         
\usepackage{graphicx}        
\usepackage{multicol}        
\usepackage[bottom]{footmisc}

\makeindex             

\begin{document}

\title*{Pulsations of rapidly rotating evolved stars}
\author{Ouazzani, R-M., Roxburgh,I.W., Dupret,M-A.}
\institute{Ouazzani, R-M. \at LESIA, Observatoire de Paris, Meudon, France, \email{rhita-maria.ouazzani@obspm.fr}
\and Roxburgh,I.W. \at Astronomy Unit, Queen Mary University of London, UK, 
\and Dupret,M-A. \at Institut d'astrophysique et de g\'eophysique, Universit\'e de Li\`ege, Belgique}
%
%
\maketitle

\vspace*{-4cm}
\abstract{A new two dimensional non-perturbative code to compute accurate oscillation modes of rapidly rotating stars is presented. The 2D calculations fully take into account the centrifugal distorsion of the star while the non perturbative method includes the full influence of the Coriolis acceleration. This 2D non-perturbative code is used to study pulsational spectra of highly distorted evolved models of stars. 2D models of stars are obtained by a self consistent method which distorts spherically averaged stellar models a posteriori. We are also able to compute gravito-acoustic modes for the first time in rapidly rotating stars. We present the dynamics of pulsation modes in such models, and show regularities in their frequency spectra. 
}

\section{Introduction}
Fast rotation is known to have a strong impact on stellar structure and pulsations: centrifugal distortion breaks the thermal equilibrium in a star and induces transport of chemical elements and angular momentum through large scale currents and shear turbulence, the centrifugal force distorts the pulsations cavity and the Coriolis force modifies the dynamics of modes.
For many stars such as, for instance, $\delta$ Scuti stars and Be stars, rotation has to be taken into account in order to use seismology as a tool to probe their interior. 
So far the impact of fast rotation on observational frequency spectra is poorly understood, and some questions still need to be answered. What are the modes corresponding to each observed frequency? Which cavity inside the star do they probe? How can the physics inside this cavity can be extracted? All these questions call for realistic modelling.
\vspace*{-0.5cm}

\section{The model}
\subsection{The 2D structure}

At the present time there are no fully two-dimensional stellar evolution models. Some progress is being made in the context of the ESTER project (for \textit{Evolution STEllaire en Rotation} \cite{Rieutord2007}), but such models are not yet available.
An alternative is to calculate spherically averaged stellar evolution models and then to distort the models \textit{a posteriori} using a self-consistent method. We present here such an approach based on a method established by \cite{Roxburgh2006}. It consists of building the acoustic structure of a rotating star in hydrostatic equilibrium. Poisson's equation together with the curl of hydrostatic equilibrium equation are alternatively integrated, taking into account the centrifugal force. This is done through an iterative scheme which starts with spherical profiles for the density and, given the angular velocity,  leads to 2D profiles for density and gravitational potential. The pressure and $\Gamma_1$ profiles are then found by solving the hydrostatic equilibrium and the equation of state.

\subsection{2D non perturbative pulsations}
For the computation of pulsations we only require the hydrostatic and dynamical structure of the star and not the thermal structure. We compute the oscillation modes as the adiabatic response of the structure to small perturbations -- i.e. of the density, pressure, gravitational potential and velocity field -- using the Eulerian formalism. 

The code developed for this purpose is named ACOR --Adiabatic Computations of Oscillations including Rotation--. It consists in solving the hydrodynamics equations perturbed by Eulerian fluctuations, by direct integration of the two-dimensional problem.
The numerical method is based on a spectral multi-domain method which expands the angular dependence of pulsation modes in spherical harmonics series, and whose radial treatment is particularly well adapted to the behaviour of equilibrium quantities in evolved models at the interface of convective and radiative regions, and at the stellar surface. The radial differentiation is made by means of a sophisticated finite difference method which is accurate up to the fifth order in terms of the radial resolution.

This code has been validated by comparison with the results of \cite{Reese2006} for polytropic models. The agreement between the two codes is found excellent.

\section{Results}
We apply this modelling process to a 2 M$_{\odot}$ non barytropic, differentially rotating star, with a surface angular velocity of $80 \%  \Omega_k$, where $\Omega_k$ is the Keplerian critical angular  velocity ($\Omega_k=\sqrt{\rm G M / R_e^3}$, $R_e$ is the equatorial radius). This model star has been evolved from an initial composition of $X_c=0.72$ to $X_c=0.35$. The angular velocity varies with radius $\Omega(r)$ with $\Omega_{\rm center}= 3\Omega_{\rm surface}$.
\begin{figure}[t!]
\begin{center}
\includegraphics[scale=0.7]{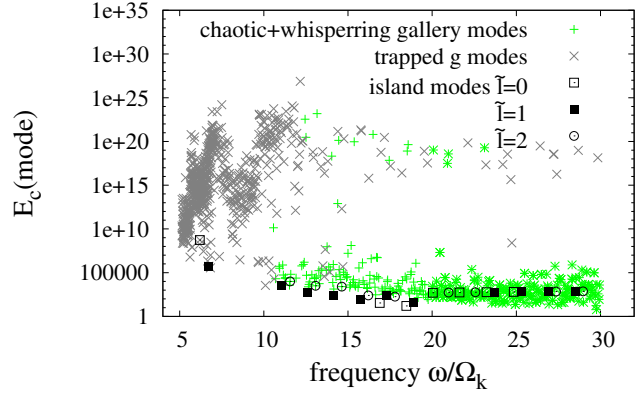}
\caption{\label{freq-vs-Ec} Frequency vs Energy diagram of the pulsation modes computed for a 2D model of 2 M$_\odot$ evolved star.}\vspace*{-0.5cm}
\end{center}
\end{figure}
Depending on the region of the frequency spectrum explored by our computations, five classes are found. The frequency vs energy distribution of these five classes of modes are presented in Fig.\ref{freq-vs-Ec}. 
The kinetic energy is the Eulerian equivalent of the inertia of modes, and is given by: $E_c(mode)\,  = \, \int_{\star} dm \mid v'(r)\mid^2 \,  /  \, \mid v'(r=R)\mid^2 $.
As in the non rotating case, modes trapped in the core have higher inertia than those trapped in the envelope. Therefore, this diagram gives access to the nature -- i.e. gravity, pressure or gravito-acoustic-- of the computed modes.

At low frequency, represented by grey dots in Fig.\ref{freq-vs-Ec}, we mostly find modes trapped in the region of sharp variation of the mean molecular weight above the convective core, characterised by a sharp feature in the  Brunt-V\"ais\"al\"a frequency. These modes are counterparts of high order g-modes in the rapidly rotating case. They have negligible amplitude in outer layers, and we expect them to undergo radiative damping, therefore, they should not be detected observationally. However the spectrum is overcrowded by this type of mode at low frequency, which makes the detection of modes trapped in the envelope in this frequency range very difficult in synthetic spectra.

At higher frequency --green in Fig.\ref{freq-vs-Ec}--, we find whispering gallery modes, chaotic modes and island modes such as those computed in homogeneous models of stars by \cite{Lignieres2006}.
Whispering gallery modes are the counterpart of high degree acoustic modes in the rapidly rotating case. They probe the outer layers of the star, but as they show very low visibility factors, they might not be detected. 

Chaotic modes --also green in Fig.\ref{freq-vs-Ec}-- have no counterparts in the non rotating case. They are found only in very fast rotating models of stars. Their spatial distribution does not present a simple symmetry. They have significant amplitude in the whole stellar interior. The lack of symmetry should result in a low cancellation factor. Therefore, these modes are expected to be detected observationally. 

\begin{figure}[t!]
\begin{center}
\hspace*{-1cm}\includegraphics[scale=0.57]{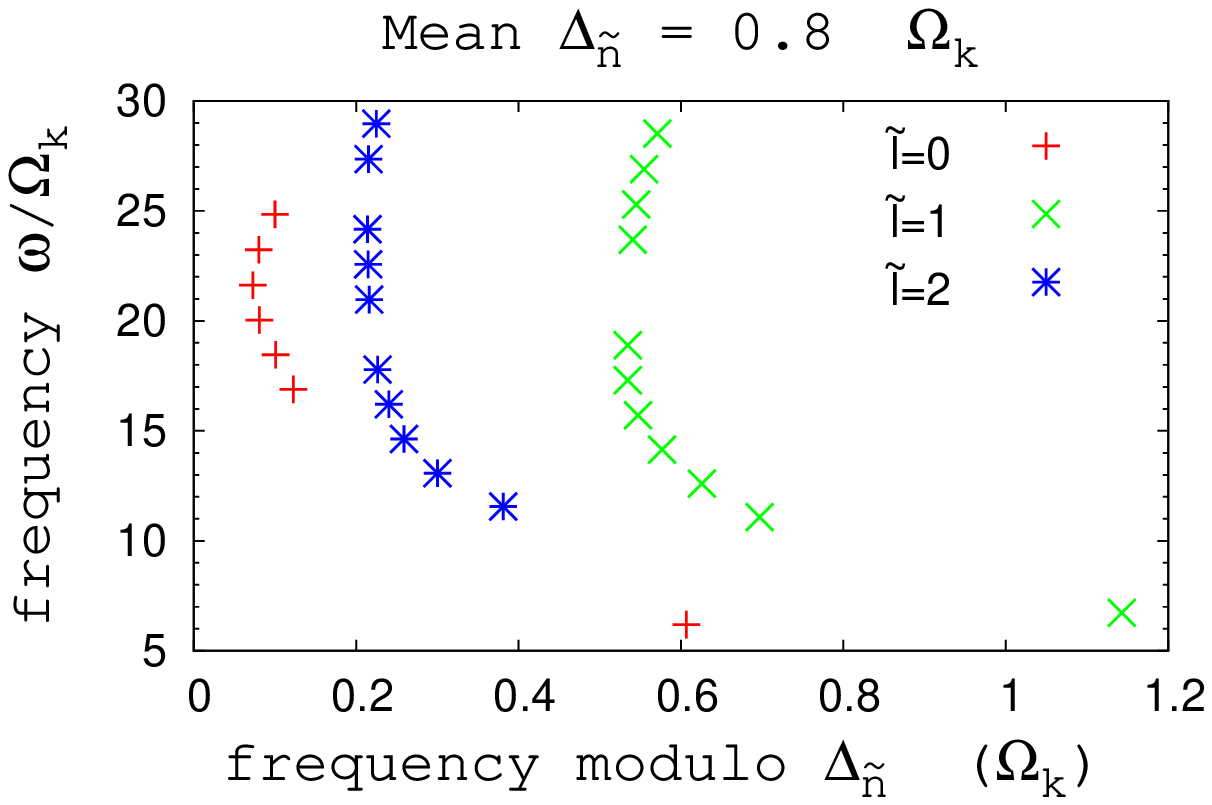}
\hspace*{0.5cm}\includegraphics[scale=0.23]{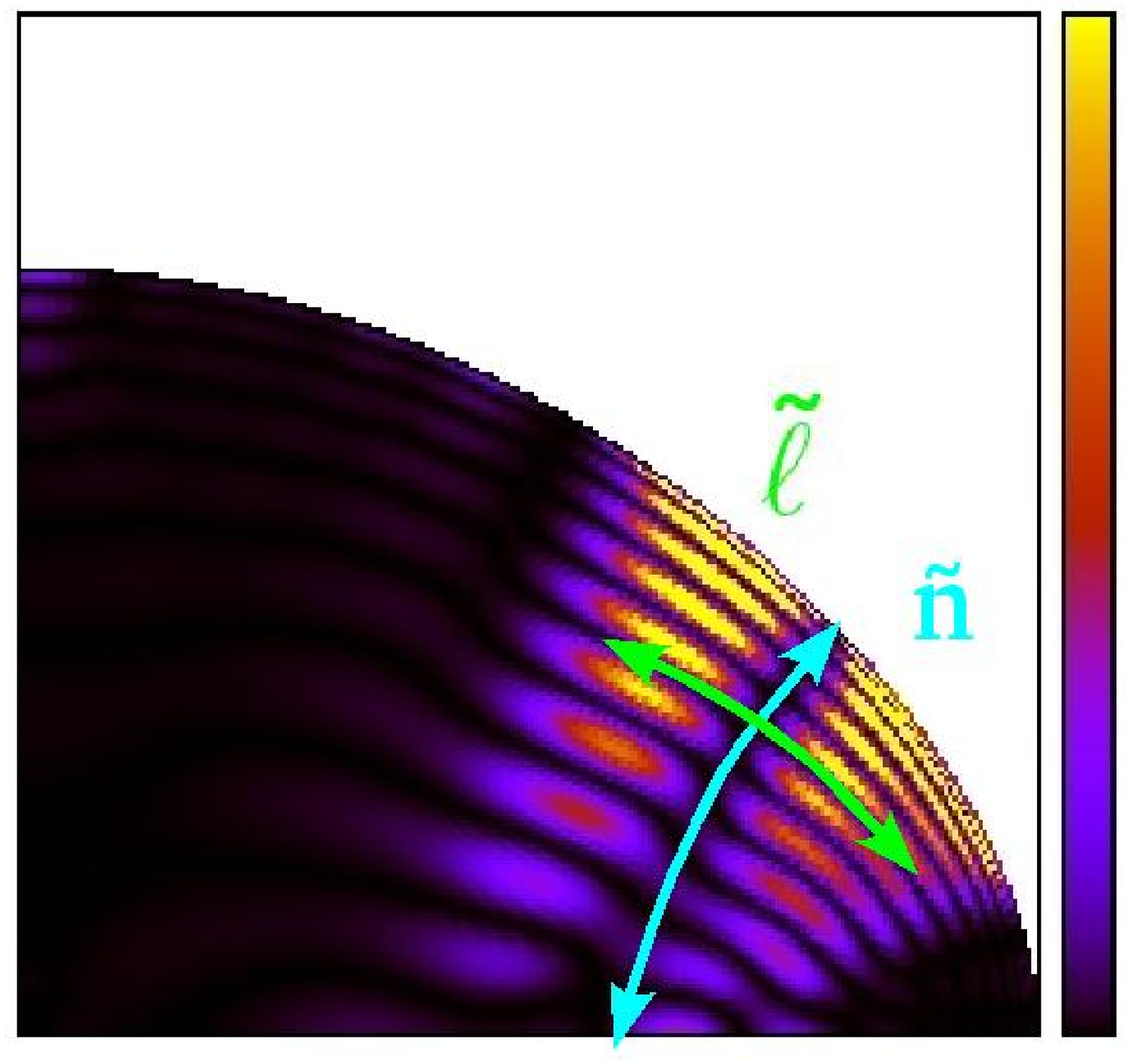}
\caption{\label{fig_Ech} \textbf{Left:} Echelle diagram for axisymmetric island modes. \textbf{Right:} Kinetic energy distribution of an island mode in a meridional plane.}\vspace*{-0.5cm}
\end{center}
\end{figure}

Island modes --black dots in Fig.\ref{freq-vs-Ec}-- are the counterparts of low degree acoustic modes in the rapidly rotating case. They probe the  outer layers of the star, and present good visibility factors. Therefore, they should be easily detected observationally. They are characterised by two quantum numbers $\tilde{\ell}$ and $\tilde{n}$ --the number of nodes along and perpendicular to the ray path, see Fig. \ref{fig_Ech}--. 

Some island modes --the most energetic island modes in Fig.\ref{freq-vs-Ec} right panel--  are gravito-acoustic modes --mixed modes-- such as for example the $\tilde{\ell}=0$ and the $\tilde{\ell}=1$ island modes which frequency is around $6.5 \Omega_k$. These modes have significant amplitude in the stellar core as well as in the envelope. Therefore these modes hold very interesting seismic diagnostics, they should have a good visibility factor and carry information about both the core and the envelope. These modes could provide valuable knowledge on the internal physics of rapidly rotating evolved stars including their differential rotation profile.

In Fig. \ref{fig_Ech} (left) is given the Echelle diagram of axisymmetric modes computed for the model mentionned earlier. This echelle diagram is very similar to echelle diagrams of acoustic modes in non-rotating stars, but it is built on different quantum numbers (see right panel of Fig. \ref{fig_Ech}). This illustrates that with the appropriate definitons for the quantum numbers island modes spectrum is regular. This new kind of large separation should then provide a seismic diagnostic which is valid for rapid rotators \cite{Pasek2011}. Moreover, in the low frequency regime, the spectrum is less regular, and the ridges adopt a particular curvature. In the non rotating case, this curvature is attributed to the mixture of acoustic and gravity modes. In this case, where the star is at the middle of its life on the main sequence and is fastly rotating, one can wonder how rotation impacts this deviation from regularity. 



\begin{acknowledgement}
The authors would like to thank The SOC and the LOC for a very lively, fruitful and enjoyable meeting.
IWR gratefully acknowledges support from STFC under grant PP/E001815/1 
\end{acknowledgement}

\end{document}